\documentclass[reprint,aps,prb,amsmath,amssymb]{revtex4}

\usepackage[dvipdfmx]{graphicx}
\usepackage[dvipdfmx]{color}
\usepackage{url}
\usepackage{dcolumn}
\usepackage{bm}
\usepackage[dvipdfmx,bookmarkstype=toc=true,colorlinks=true,urlcolor=blue,%
      linkcolor=blue,citecolor=blue,linktocpage=true,bookmarks=true]{hyperref}
\usepackage{xspace}
\usepackage{comment}

\begin{document}

\title{Superconductivity in an electron band just above the Fermi level: possible route to BCS-BEC superconductivity}

\author{K.~Okazaki$^{1,*}$}
\author{Y.~Ito$^{1}$}
\author{Y.~Ota$^{1}$} 
\author{Y.~Kotani$^{1,\dag}$}
\author{T.~Shimojima$^{1,\ddag}$} 
\author{T.~Kiss$^{1,\S}$}
\author{S.~Watanabe$^{2}$}
\author{C.~-T.~Chen$^{3}$} 
\author{S.~Niitaka$^{4,5}$} 
\author{T.~Hanaguri$^{4,5}$} 
\author{H.~Takagi$^{4,5}$}
\author{A.~Chainani$^{6,7}$}
\author{S.~Shin$^{1,5,6,8}$}
\affiliation{
$^{1}$Institute for Solid State Physics (ISSP), University of Tokyo, Kashiwa, Chiba 277-8581, Japan\\
$^{2}$Research Institute for Science and Technology, Tokyo University of Science, Chiba 278-8510, Japan\\
$^{3}$Beijing Center for Crystal R\&D, Chinese Academy of Science (CAS), Zhongguancun, Beijing 100190, China\\
$^{4}$RIKEN Advanced Science Institute, 2-1, Hirosawa, Wako, Saitama 351-0198, Japan \\
$^{5}$TRIP, JST, Chiyoda-ku, Tokyo 102-0075, Japan\\
$^{6}$RIKEN SPring-8 Center, Sayo-gun, Hyogo 679-5148, Japan\\
$^{7}$Department of Physics, Tohoku University, Aramaki, Aoba-ku, Sendai 980-8578 Japan\\
$^{8}$CREST, JST, Chiyoda-ku, Tokyo 102-0075, Japan\\
$^{*}$Present address: Department of Physics, University of Tokyo, Tokyo 113-0033, Japan\\
$^{\dag}$Present address: Japan Synchrotron Radiation Research Institute (JASRI/SPring-8), Sayo, Hyogo 679-5198, Japan\\
$^{\ddag}$Present address: Department of Applied Physics, University of Tokyo, Tokyo 113-8656, Japan\\
$^{\S}$Present address: Graduate School of Engineering Science, Osaka University, Osaka 560-8531, Japan\\
Correspondence and requests for materials should be addressed to K.O. (okazaki@wyvern.phys.s.u-tokyo.ac.jp) or S.S. (shin@issp.u-tokyo.ac.jp).
}

\begin{abstract}
Conventional superconductivity follows Bardeen-Cooper-Schrieffer(BCS) theory of electrons-pairing in momentum-space, while superfluidity is the Bose-Einstein condensation(BEC) of atoms paired in real-space. These properties of solid metals and ultra-cold gases, respectively, are connected by the BCS-BEC crossover.
Here we investigate the band dispersions in FeTe$_{0.6}$Se$_{0.4}$($T_c$ = 14.5 K $\sim$ 1.2 meV) in an accessible range below and above the Fermi level($E_F$) using ultra-high resolution laser angle-resolved photoemission spectroscopy. We uncover an electron band lying just 0.7 meV ($\sim$ 8 K) above $E_F$ at the $\Gamma$-point, 
which shows a sharp superconducting coherence peak with gap formation below $T_c$.
The estimated superconducting gap $\Delta$ and Fermi energy $\epsilon_F$ 
indicate composite superconductivity in an iron-based superconductor, consisting of strong-coupling BEC in the electron band and weak-coupling BCS-like superconductivity in the hole band. The study identifies the possible route to BCS-BEC superconductivity.
\end{abstract}

\maketitle

The iron-based high-temperature superconductors~\cite{Kamihara2008JACS} possess multi-band hole and electron Fermi surfaces (FSs) which have motivated the role of FS nesting, spin- and orbital-fluctuations, or a combination of these mechanisms for understanding their properties~\cite{Mazin2008PRL,Kuroki2008PRL,Kontani2010PRL}. Electronic structure studies have concentrated on band dispersions in the occupied states lying below the Fermi level ($E_F$)~\cite{Lu2008Nature,Zabolotnyy2009Nature,Lubashevsky2012NP}, which open up a superconducting gap ($\Delta$) below $T_c$.
However, the observation of a pseudogap in the normal phase of the high-$T_c$ copper oxide superconductors~\cite{Ding1996Nature,Loeser1996Scisence} identified a remarkable challenge in our understanding of superconductivity and its origin is still extensively debated. In particular, does the pseudogap represent preformed Cooper pairs, or does it reflect another ground state competing with superconductivity~\cite{Lee2007Nature,Yang2008Nature,Kondo2009Nature}? More recently, a pseudogap was also found in the strongly interacting ultracold Fermi gas above the superfluid condensation temperature~\cite{Gaebler2010NP}. These fascinating results suggest the important role of pairing in the normal phase, above the onset of superfluidity/superconductivity in strongly interacting systems. The existence of a pseudogap state has been discussed in the scheme of BCS-BEC crossover~\cite{Chen2005PR}. Very interestingly, the study on ultracold Fermi gases reported a pseudogap for a system with 
$\Delta/\epsilon_F$ $\sim$ 1~\cite{Gaebler2010NP} while the optimally doped copper oxide Bi2212 shows a 
$\Delta/\epsilon_F$ $\sim$ 0.1~\cite{Lubashevsky2012NP}. In a recent study on the Fe(Te,Se) superconductor, it was concluded that the system was in the BCS-BEC crossover regime with a $\Delta$/$\epsilon_F$ $\sim$ 0.5 associated with a hole band centered at the $\Gamma$-point~\cite{Lubashevsky2012NP}. The obtained band dispersions were similar to prior studies of the normal phase~\cite{Chen2010PRB,Tamai2010PRL}, while $\Delta$ was much smaller than an earlier report which claimed strong coupling superconductivity~\cite{Nakayama2010PRL}. Although a pseudogap is expected based on theories of the BCS-BEC crossover~\cite{Eagles1969PR,Leggett1980,Randeria1989PRL,S'adeMelo1993PRL}, no study to date has identified a pseudogap in the momentum resolved electronic structure of Fe(Te,Se).
Further, a tuning of the BCS to BEC regimes across the Feshbach resonance using a magnetic field which eventually leads to a change in sign of the chemical potential for fermions~\cite{S'adeMelo1993PRL}, is well-known in ultracold atomic gases~\cite{Timmermans2001PLA,Regal2004PRL}. 
However, a similar control of the chemical potential in a solid is not possible, and consequently, the BEC superconductivity in a solid has eluded experiments. In the following, we report 
experiments indicative of composite superconductivity in an iron-based superconductor: Cooper pairing in a hole band coexisting with Bose-Einstein condensation in an electron band.

\section*{Results}
\subsection*{Electron band lying just above $E_F$}
Figure 1(a) shows an intensity plot of $E$ vs. $k$ (energy vs. momentum) measured at 25 K ($> T_c$ = 14.5 K) along $\Gamma$-$X$ line in the Brillouin zone of FeTe$_{0.6}$Se$_{0.4}$ after dividing by the Fermi-Dirac (FD) function broadened with the Gaussian corresponding to the experimental energy resolution. We employed three different methods to determine the band dispersions: a second derivative map with respect to energy, fitting to the energy distribution curves (EDCs), and fitting to the momentum distribution curves (MDCs)~(see Supplementary Information). Three hole bands can be clearly recognized in the second derivative map.   
Band-structure calculations based on density functional theory (DFT) were carried out for the parent FeTe and are overlaid as solid lines in Fig. 1(a) after a suitable energy shift and rescaling which are ascribed to renormalization effects, as is known from earlier work~\cite{Chen2010PRB,Tamai2010PRL,Nakayama2010PRL,Lubashevsky2012NP}. The calculation details and complete band structure are discussed in the Supplementary Information and the energy shifts and rescaling are listed in Table I. The calculated band structure and orbital characters are also consistent with known results~\cite{Subedi2008PRB,Miyake2010JPSJ} and were confirmed by measuring the linear polarization dependence of spectral intensities~(see Supplementary Information). However, in contrast to the DFT calculations which predict existence of three hole FSs around the $\Gamma$ point, we find that the band top of the two dominantly $xz/yz$-orbital derived bands are located around 15 meV below $E_F$, i.e., these bands sink below $E_F$, and only one hole band originating in the $x^2-y^2$ orbital crosses $E_F$. From the degeneracy of the two $xz/yz$ bands, we conclude that $k_z$ $\sim$ 0 in the reduced Brillouin zone for the present laser ARPES measurements. 

The open circles in Fig. 1(a) show band dispersions deduced from the peak positions of the second derivative spectra shown in Fig. S3. The second derivative map after dividing by the FD function shown in Fig. S3(b) clearly shows that the dispersion around the $\Gamma$ point is electron-like. The origin of this electronic dispersion is presumably another dominantly $xz/yz$-orbital derived band, which is located just above $E_F$ for the DFT results for the parent FeTe\cite{note_DFT}. For checking the dispersions above $E_F$, Fig. 1(b) shows the band dispersions in a narrow energy window near $E_F$, and band dispersions deduced from fits to the EDCs are overlaid. 
The fits to the EDCs, obtained after dividing by the FD function, are shown in Fig. 1(c). It is clear that there are two bands above $E_F$ at the $\Gamma$-point, which get merged around $k$ $\sim$ 0.07 {\AA} and then again separate out into two bands around $k$ $\sim$ 0.1 {\AA}. Figure 1(d) shows the fits with the component Lorentzian functions for these three cases. 
We also performed measurements with another sample at higher temperatures of 35 K and 50 K in addition to 25 K as shown in Fig. S7. The peak positions are consistent with those shown in Fig. 1.

In the occupied states below $E_F$, the degenerate band top of the $xz/yz$ hole bands are positioned at $\sim$ 15 meV below $E_F$. The band top of the $E_F$-crossing $x^2-y^2$ band is located at least above $\sim$ 6.5 meV from $E_F$ at the $\Gamma$-point. Most interestingly, we do find the expected electron band existing just above $E_F$ at the $\Gamma$-point, with the band bottom located at $\sim$ 0.7 $\pm$ 0.2 meV above $E_F$ (Fig. 1(b)). This electron band has been missed in all earlier studies of the momentum resolved electronic structure of Fe(Te,Se)~\cite{Chen2010PRB,Tamai2010PRL,Nakayama2010PRL,Lubashevsky2012NP}. 
We note that the $x^2-y^2$ hole band and the electron-like band just above $E_F$ may be hybridized due to spin-orbit interactions~\cite{note_SOI}, for example, and result in a wing-shaped dispersion. However, even if these two bands are hybridized and merge to a single band, this does not affect our conclusions. Since the nature of the conducting carriers being electron-like or hole-like is determined by the gradient of the band dispersion ($\partial E/\partial k$), the carriers at $k_F$ and the thermally-excited carriers at the $\Gamma$ point will be hole-like and electron-like, respectively.  
Also, the details of electron band at the higher energy region above $E_F$ are not relevant to superconductivity. Only the positions of the top of hole band and the bottom of the electron band are important, and they can be evaluated rather clearly from the MDCs, of which line shape is not affect by dividing by the FD function~(see Supplementary Information).

\subsection*{Sharp superconducting coherence peak in the electron band just above $E_F$}
Figures 2(a) and 2(b) show the energy distribution curves (EDCs) after dividing by the FD functions corresponding to each temperature along $\Gamma$-$X$ line at $T$ = 25 K (above $T_c$) and at $T$ = 2.5 K (below $T_c$), respectively. The open circles in Fig. 2(a) mark the normal-state band dispersions obtained from the second derivative spectra shown in Fig. S3. In Fig. 2(b), we can clearly see that the superconducting coherence peaks emerge below $T_c$ for the hole band. The small circles in Fig. 2(b) mark the positions of the coherence peaks, and they are plotted in the enlarged scale in Fig. S10. Figures 2(c) and 2(d) show the EDCs above $T_c$ (25 K) and below $T_c$ (2.5 K) at $k = k_F$ and $k \sim \Gamma$, respectively. We can see that the electron band just above $E_F$ at the $\Gamma$ point also shows a sharp superconducting coherence peak, although this band does not cross the $E_F$ in the normal state. The solid lines are fits to the BCS spectral function $A_\mathrm{BCS}(k,\omega)$\cite{Matsui2003PRL,Shimojima2011Science,Okazaki2012Science}, which can be expressed as
\[
A_\mathrm{BCS}(k,\omega) = \frac{1}{\pi}\left\{\frac{|u_k|^2\Gamma}{\left(\omega-E_k\right)^2+\Gamma^2}+\frac{|v_k|^2\Gamma}{\left(\omega+E_k\right)^2+\Gamma^2}\right\},
\]
where $E_k$ and $|u_k|^2$, $|v_k|^2$ are the quasiparticle energy and the coherence factors of Bogoliubov quasiparticles (BQPs), respectively. Using the normal-state dispersion $\epsilon_k$ with respect to the chemical potential $\mu$ and the SC gap $\Delta(k)$, $E_k$ can be expressed 
\[
E_k = \sqrt{(\epsilon_k-\mu)^2 + |\Delta(k)|^2}
\]
and 
\[
|u_k|^2 = 1 - |v_k|^2 = \frac{1}{2}\left(1+\frac{\epsilon_k}{E_k}\right),
\]
respectively. From the fits to the data, we estimate the Bogoliubov quasiparticle energy ($E_k = \sqrt{\epsilon_k^2 + |\Delta(k)|^2}$) of the hole band to be 2.3 meV (= $\Delta(k)$, because $\epsilon_k$ = 0 at $k$ = $k_F$ of this band) and of the electron band to be 1.3 meV, respectively. From the value of $\epsilon_k$ $\sim$ 0.7 meV, $\Delta(k)$ is estimated to be $\sim$ 1.1 meV for the electron band. This indicates different pairing strengths and reduced gap values 2$\Delta$/$k_BT_c$ for the electron and hole bands. 
In addition, $\Delta$/$\epsilon_F$ for the hole band is estimated to be $\sim$ 0.3, corresponding to a relatively weak coupling. On the other hand, 
since the energy position of the electron band is just 0.7 meV ($\sim$ 8 K) above $E_F$, it means that its occupancy in the normal state will strongly depend on temperature. Accordingly, the exact value of $\epsilon_F$ of the electron band cannot be described in the usual way.
If we regard $T_c$ as a measure of $\epsilon_F$, based on the fact that $T_c$ is the lowest temperature representing the normal state ($T_c$ = 14.5 K $\sim$ 1.2 meV), we obtain $\Delta$/$\epsilon_F$ $\geq$ 1, 
indicative of the strong coupling limit. This estimation may seem to be fairly rough. However, if $\epsilon_F$ equals to $\Delta$, the bottom of the electron band should be located at $E$ = -$\Delta$ below $E_F$. Hence, we can say at least $\Delta$/$\epsilon_F$ $\geq$ 1. Thus, the electron band with a smaller $\Delta$ is actually in the strong-coupling regime. This represents the condition of an electron band with only a small number of carriers, but with a strong pairing interaction and a finite $\Delta$ exists for this band. On the otherhand, the hole band with a larger $\Delta$ lies in the relatively weaker-coupling regime.
It is suggestive of Cooper pairing for the hole band and Boson condensation for the electron band.
We note that even for the strong-coupling electron band, we have used a BCS spectral function to estimate the value of $\Delta$. This is not a problem as the obtained value of $\Delta$ represent the lower bound of $\Delta$, because a smaller value of $\epsilon_k$-$\mu$ for the BEC regime will give a larger value of $\Delta$\cite{Gaebler2010NP,Lubashevsky2012NP}.
 
Figures 2(e) and 2(f) show the intensity maps of the spectra above and below $T_c$, respectively, after dividing by the corresponding FD functions. The open circles are the same as in Fig. 2(a) and the solid line is a fitting result to the open circles using a polynomial function, representing the normal-state dispersion $\epsilon_k$~\cite{note_dispersion}. The solid lines in Fig. 2(f) are the BQP dispersions using $\epsilon_k$ in Fig. 2(e) and the $\Delta_k$ values obtained above for the electron and hole bands. Colors of the lines corresponds to the amplitude of the coherence factors $|u_k|^2$ above $E_F$ and $|v_k|^2$ below $E_F$. The red and blue regions correspond to the higher and lower values, respectively. It is noted that the BQP dispersion does not cross the brightest intensity around the $\Gamma$ point above $E_F$. This is attributed to the tail of the hole-band top, which is also positioned at the $\Gamma$-point. The normal-state and BQP dispersions have been plotted in the same panel of Fig. 2(g). The open circles plotted in Fig. 2(g) correspond to the dispersion of the coherence peaks at $T$ = 2.5 K shown in Fig. 2(b). The gray-scale density of the BQP dispersion corresponds to the amplitude of coherence factors. 
It is interesting to note that the BQP dispersions merge for the electron and hole bands, indicative of a composite BCS-BEC superconductivity. The results indicate that irrespective of weak or strong coupling
, both the hole and electron bands in the superconducting state exhibit Bogoliubov quasiparticle dispersions due to particle-hole mixing~\cite{Campuzano1996PRB}.  However, the superconductivity in the electron band can be expected to be very sensitive to the occupancy of the electron band with Se substitution for Te, as well as pressure/strain. This possibly explains the reported large variation in $T_c$ with pressure for FeSe ($T_c$ = 8.5-36.7 K)~\cite{Medvedev2009NM}.

\section*{Discussion} 
Another difference can be recognized between the weak-coupling hole band and the strong-coupling electron band. Figures 3(a) and 3(b) show the temperature dependence of EDCs at $k_F$ for the hole band and $k \sim \Gamma$ for the electron band, respectively. The black solid lines indicate the fitting results using the BCS spectral function. The estimated SC-gap sizes are shown in Fig. 3(c). The existence of the pseudogap only for the hole band is clearer from the symmetrized EDCs (Fig. S11) or the FD-divided EDCs (Fig. S12).
The temperature dependence of the gap opening indicates another important difference for the weaker-coupling hole band compared to the strong-coupling electron band. 
A pseudogap behavior can be recognized for the weaker coupling hole band in the spectra above $T_c$~\cite{note_pseudogap}. However, in strong contrast to the currently available BCS-BEC crossover theory~\cite{S'adeMelo1993PRL} which predict existence of a pseudogap in the BEC strong coupling regime, the electron band does not show a pseudogap above $T_c$~\cite{note_BQPcontinuation}. Thus, we find a coexistence of the weak coupling and strong coupling superconductivity in the same material but with attributes not fully consistent with our present understanding of weak and strong coupling superconductivity. 
Our study identifies the required band structure for composite superconductivity, which is closely related to Dirac point dispersions, coexisting with a simple electron or hole band as schematically shown in Fig. 4.

\section*{Methods}
Single crystals of FeTe$_{0.6}$Se$_{0.4}$ were prepared by a melt-growth technique. Chemical composition of the grown crystals was determined by electron probe microanalysis (EPMA) and inductively coupled plasma (ICP) atomic emission spectrometry. Details have been described in Ref.~\onlinecite{Hanaguri2010Science}. ARPES data were collected using the laser ARPES apparatus developed at ISSP with the 6.994 eV, 6th harmonic of Nd:YVO$_4$ quasi continuous wave (q-CW, repetition rate = 120 MHz) laser and VG-Scienta HR8000 electron analyzer~\cite{Okazaki2012Science}. While this apparatus achieves the maximum energy resolution of 70 $\mu$eV, the overall energy resolution was set to $\sim$ 1.2 meV for the measurements of EDCs and MDCs near $E_F$ and 5 meV for $E$-$k$ map measurements, The angular resolution was 0.1 deg, corresponding to the momentum resolution of 0.0015 {\AA}$^{-1}$. Polarization of incident excitation laser was adjusted using a half-wave ($\lambda$/2) plate and a quarter-wave ($\lambda$/4) plate. 
The $E_F$ positions were calibrated by measuring the Fermi edge of a gold film evaporated onto the sample substrate.

\vspace{1cm}

{\bf Acknowledgements} We would like to thank K. Kuroki and R. Arita for valuable discussions and comments. This research is supported by JSPS through its FIRST Program.

\vspace{5mm}

{\bf Author Contributions}
K.O., Y.O., Y.K., T.S. and T.K. developed the laser-ARPES apparatus; S.W. and C.T.C. developed the 7-eV laser system using the nonlinear optical crystal; S.N., T.H., and H.T. grew high-quality single crystals and characterized them; K.O, Y.I. and Y.O. performed the experiments and data analysis; K.O. and A.C. wrote the manuscript; S.S. designed the project; All authors discussed the results and commented on the manuscript.

\vspace{5mm}

{\bf Additional Information}
Supplementary Information accompanies this paper at http://www.nature.com/scientifcreports.

Competing financial interests: The authors declare no competing financial interests. 


\begin{table}[htbp]
\begin{center}
{{\bf Table 1 $\mid$ Energy shifts and rescaling values required for matching the calculated dispersions with the experimental dispersions.}
\label{tableI}
}
{\tabcolsep = 6 mm
\begin{tabular}{|c|c|c|}
\hline
~      & shift    & scale \\
\hline
31 st & -130 meV & 1/15  \\
\hline
30 th & -278 meV & 1/10  \\
\hline
29 th & -110 meV & 1/4.5 \\
\hline
28 th & -110 meV  & 1/12   \\
\hline
\end{tabular}
}
\end{center}
\end{table}

\begin{figure*} [htbp]
\begin{center}
\includegraphics[width=16cm]{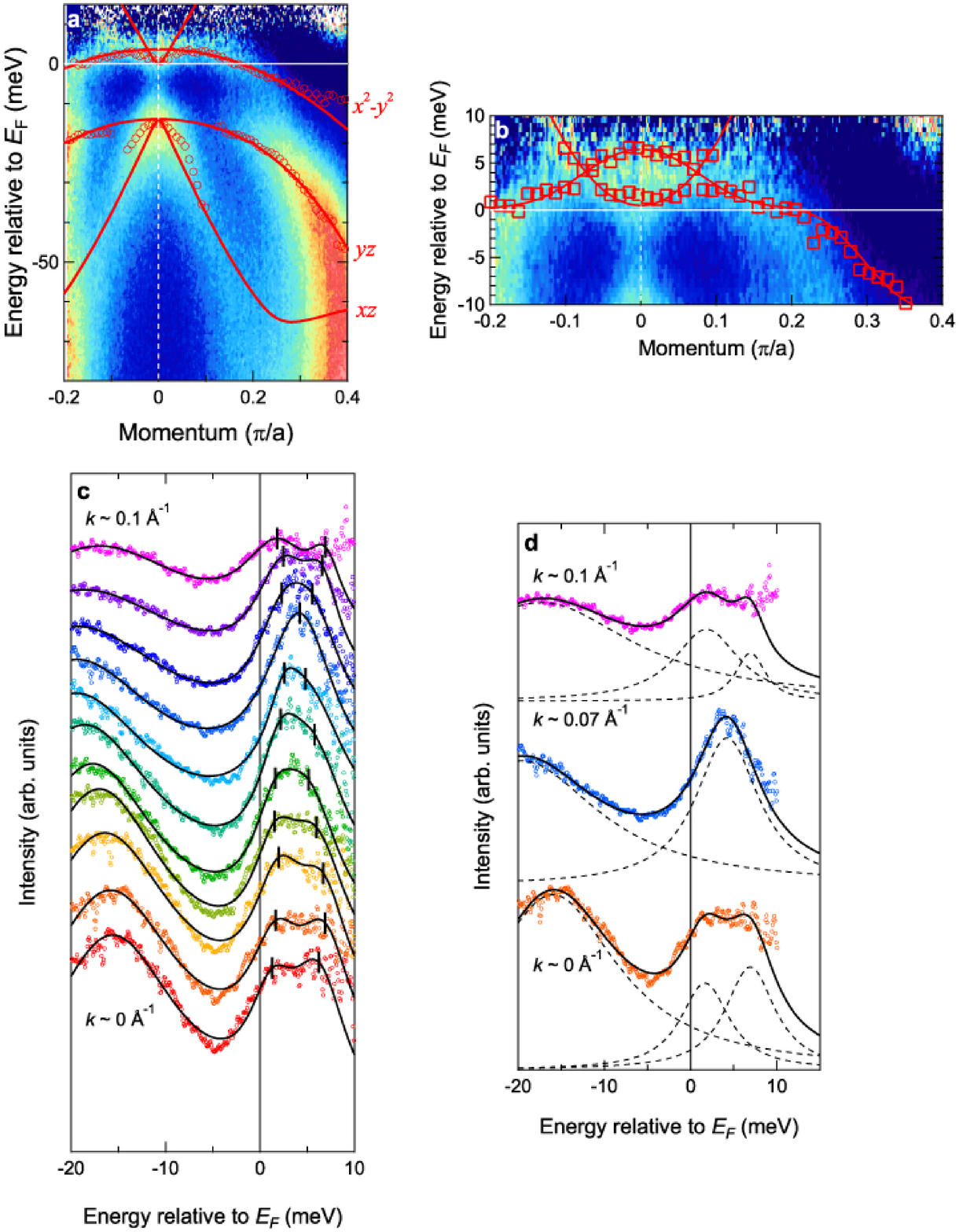} 
\end{center}
\begin{flushleft}
{{\bf Figure 1 $\mid$ Band dispersions of FeTe$_{0.6}$Se$_{0.4}$ along $\Gamma$-$X$ line measured at 25 K.} ({\bf a}) Intensity plot of $E$ vs. $k$ measured at 25 K along $\Gamma$-$X$ line. The shifted and rescaled DFT result is overlaid as solid lines. The open circles indicate band dispersions deduced from the peak positions of the second derivative spectra. ({\bf b}) Intensity plot near $E_F$. The open rectangles indicate band dispersions deduced from the fitting to the EDCs. The solid lines are dispersions by fitting to the rectangles with a polynomial function. ({\bf c}) Fitting to several cuts of EDCs after dividing by the FD function. The solid lines indicate the fitting results. The fitting functions were obtained using three Lorentzians convoluted with the Gaussian corresponding to the experimental energy resolution. ({\bf d}) Fitting results with the component Lorentzians for three cases, around the $\Gamma$ point, $k$ $\sim$ 0.07 $\AA$, and $k$ $\sim$ 0.1 $\AA$.
\label{Fig1} 
}
\end{flushleft}
\end{figure*}

\begin{figure*} [htbp]
\begin{center}
\includegraphics[height=21.5cm]{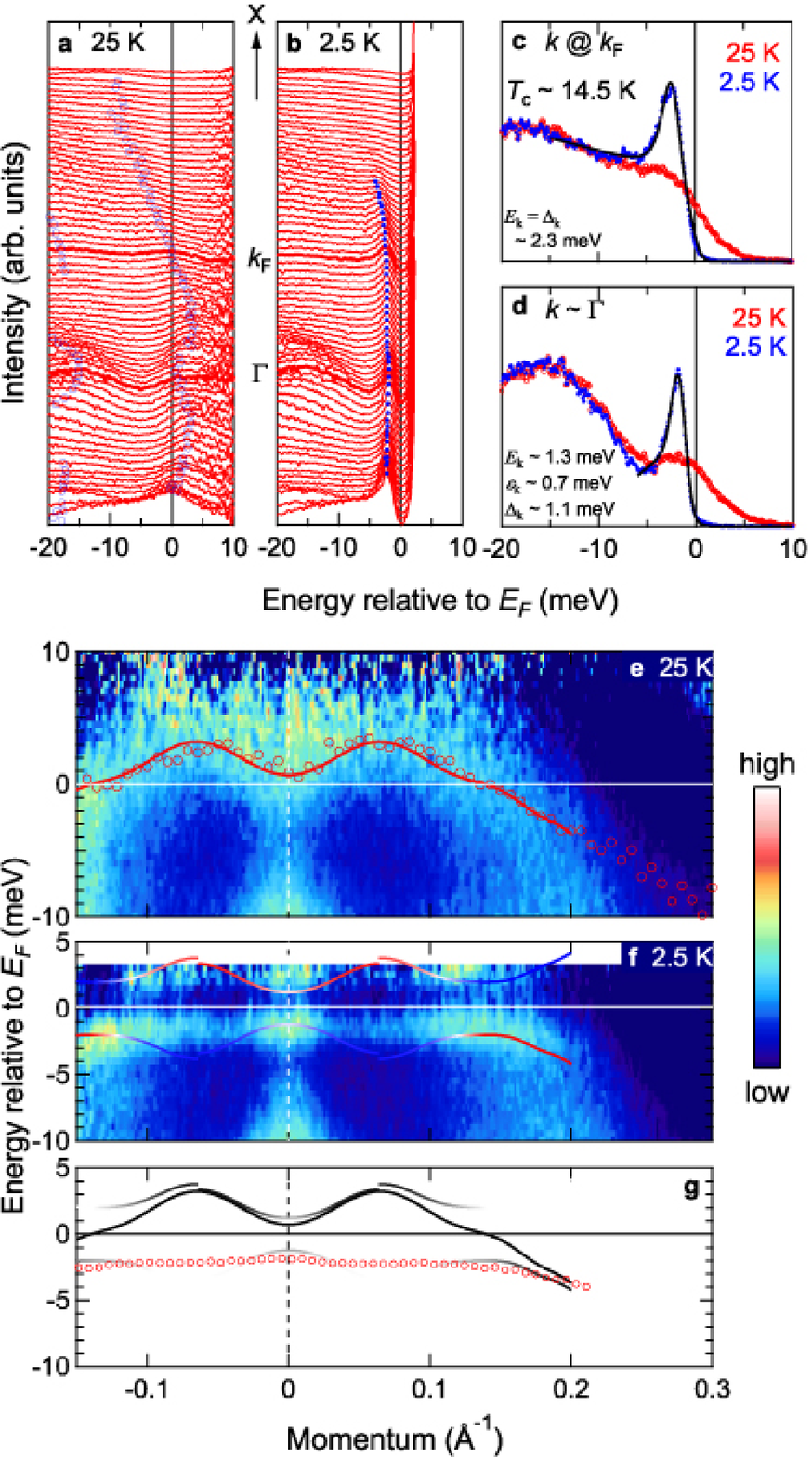} 
\end{center}
\begin{flushleft}
{{\bf Figure 2 $\mid$ Laser-ARPES spectra of FeTe$_{0.6}$Se$_{0.4}$ above and below $T_c$.} FD-divided EDCs along $\Gamma$-$X$ direction above $T_c$ ({\bf a}) and below $T_c$ ({\bf b}), respectively. EDCs above and below $T_c$ at $k = k_F$ ({\bf c}) and $k \sim \Gamma$ ({\bf d}), respectively. The solid lines are the fitting results using the BCS spectral function. Intensity plots of the spectra above  $T_c$ ({\bf e}) and below $T_c$ ({\bf f}), respectively. The solid line in panel {\bf e} is a fitting result using a polynomial function for the dispersion indicated by the open circles. The solid lines in panel {\bf f} are Bogoliubov quasiparticle (BQP) dispersions using the normal-state dispersion in panel {\bf e} and the SC gap $\Delta(k)$ = 2 meV for the hole band and  $\Delta(k)$ = 1 meV for the electron band, respectively. Colors of the lines correspond to the amplitude of coherence factors $|u_k|^2$ and $|v_k|^2$. The red and blue regions have larger and smaller coherence factors, respectively. ({\bf g}) The normal-state and BQP dispersions have been plotted in the same panel. The open circles in panel {\bf g} correspond to the dispersion of the coherence peaks at $T$ = 2.5 K in panel {\bf b}. The BQP dispersions merge for the electron and hole bands, indicative of a composite type of BCS-BEC superconductivity.
\label{Fig2} 
}
\end{flushleft}
\end{figure*}

\begin{figure*} [htbp]
\begin{center}
\includegraphics[width=16cm]{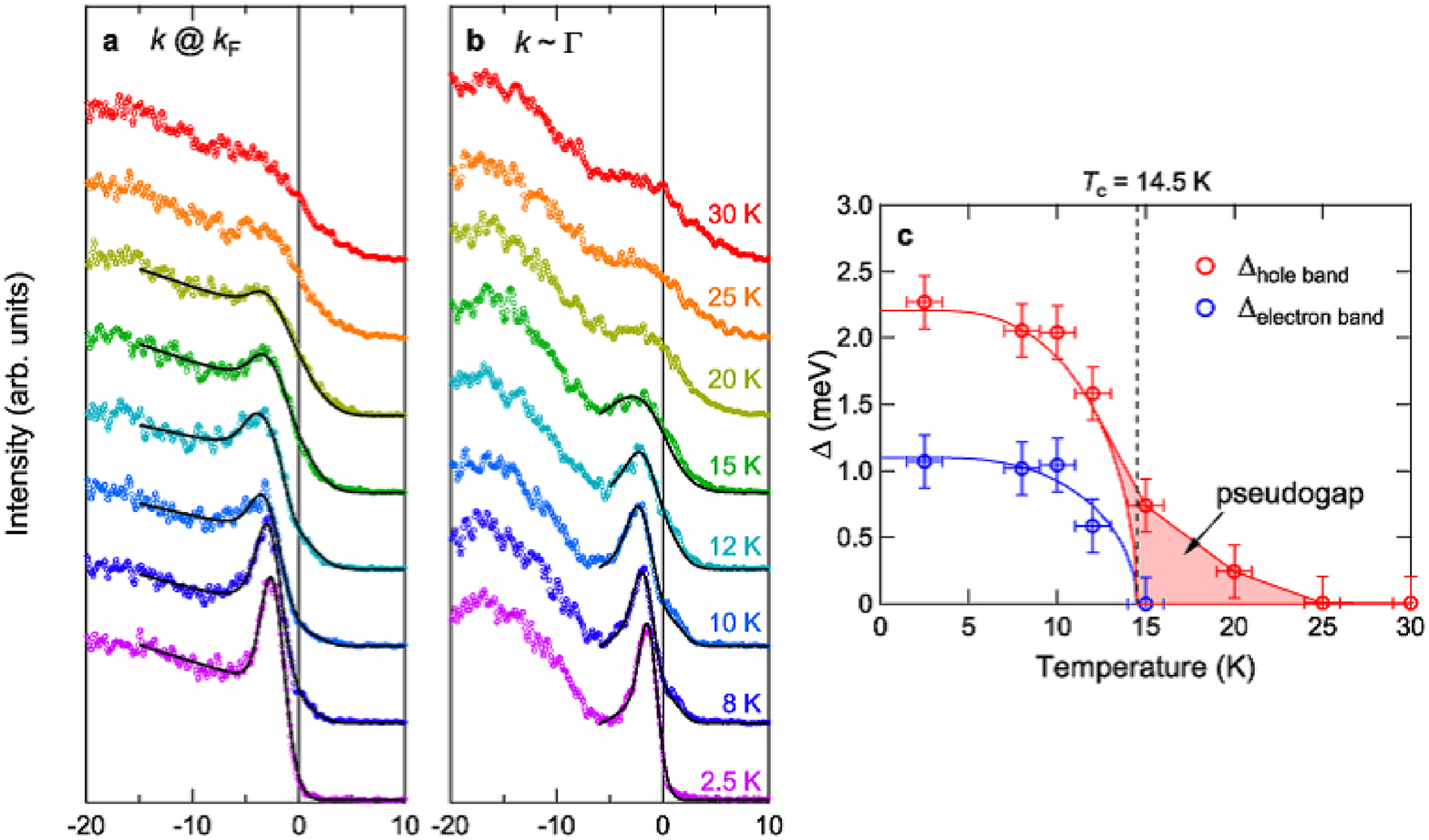} 
\end{center}
\begin{flushleft}
{{\bf Figure 3 $\mid$ Temperature dependence of EDCs at $k$ = $k_F$ and $k$ $\sim$ $\Gamma$.} Temperature dependence of raw EDCs at ({\bf a}) $k$ = $k_F$ of $x^2-y^2$ hole-like band and ({\bf b}) $k$ $\sim$ $\Gamma$ (bottom of the electron-like band), respectively. Fitting results with the BCS spectral function are indicated by black solid lines. {\bf c}, Temperature dependence of the obtained SC-gap sizes from the fitting to the BCS spectral function. Shaded area indicates pseudogap for the hole-like band.
\label{Fig3} 
}
\end{flushleft}
\end{figure*}

\begin{figure*} [htbp]
\begin{center}
\includegraphics[width=16cm]{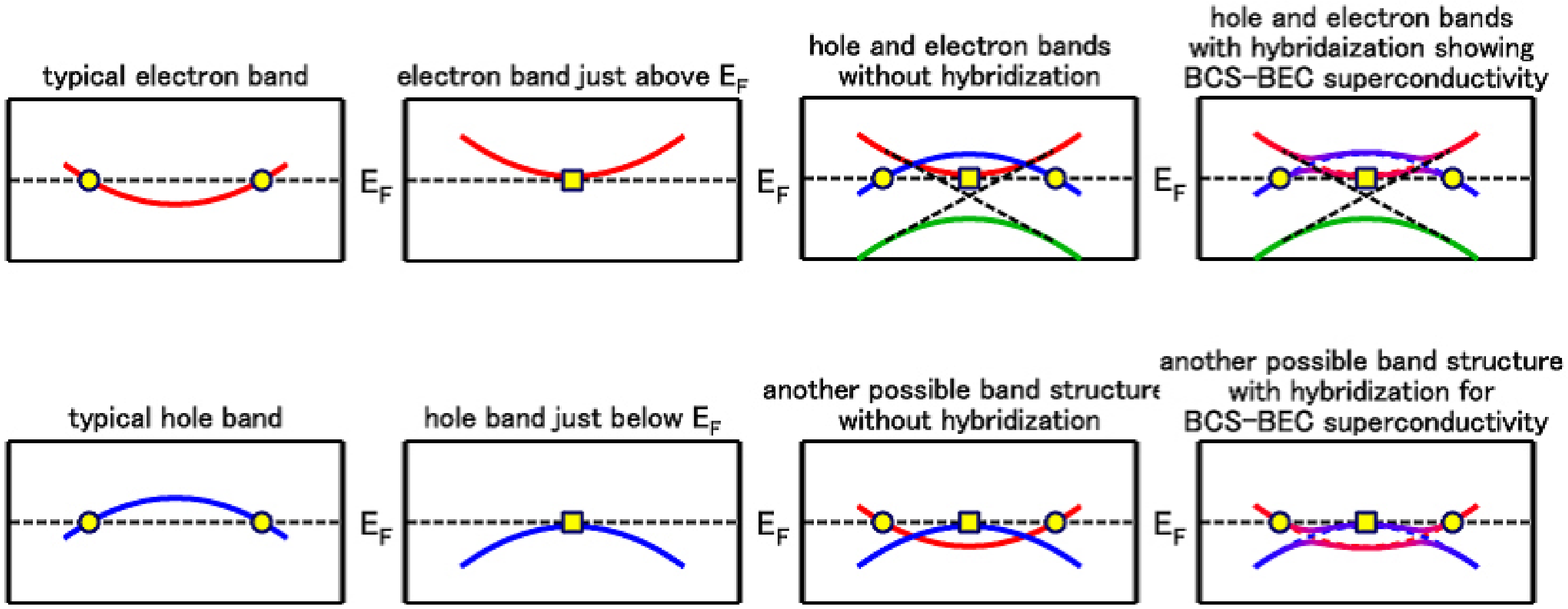} 
\end{center}
\begin{flushleft}
{{\bf Figure 4 $\mid$ Schematic band structure} showing: a typical electron band, a typical hole band, an electron band lying just above $E_F$ 
, the band structure in the absence of hybridization, and with hybridization showing composite BCS-BEC superconductivity. Similarly, a hole band lying just below $E_F$ and another possible route to BCS-BEC superconductivity. The black dashed lines show similarity to Dirac point dispersion (see also Fig. S1 for relation with band-structure calculations).
\label{Fig4} 
}
\end{flushleft}
\end{figure*}

\clearpage
\newpage


\begin{thebibliography}{10}
\bibitem{note_DFT}
{According to the reported DFT results for FeSe, the electronic disperion is
  located at $\sim$ 0.5 eV above $E_F$ at the $\Gamma$ point. Hence, we can
  expect that the electron band in the DFT regime is located at $\sim$ 0.2 eV
  above $E_F$ for FeTe$_{0.6}$Se$_{0.4}$. This is small enough to be shifted
  just above EF by a renormalization effect, because the two hole bands which
  are predicted to make Fermi surfaces by the DFT calculations get shifted
  below $E_F$.}
\bibitem{note_dispersion}
{Here, the normal state dispersions of the hole and electron bands merge to a
  single band. This means that these two bands are assumed to be hybridized at
  the band crossing.}
\bibitem{note_pseudogap}
{The exsitence of the pseudogap for the hole band can be clearly recognized
  also from the temperature-dependent symmetrized EDC (see Fig. S9)}.
\bibitem{note_BQPcontinuation}
{This observation might lead to suspect that the superconducting gap near the
  $\Gamma$ point is not originated from the electron band, but just a
  continuation from the BQP near $k_F$. However, as shown in Fig. S5, the
  observation of the superconducting coherence peaks well separately at $k =
  k_F$ and the $\Gamma$ point evidently indicate that the superconducting gap
  at the $\Gamma$ point is originated from the electron band just above $E_F$.}
\end{thebibliography}
\end{document}


\title{Supplementary Information for "Superconductivity in an electron band just above the Fermi level: possible route to BCS-BEC superconductivity"}

\author{K.~Okazaki$^{1}$}
\altaffiliation{Present address: Department of Physics, University of Tokyo, Tokyo 113-0033, Japan}
\email{okazaki@wyvern.phys.s.u-tokyo.ac.jp}
\author{Y.~Ito$^{1}$}
\author{Y.~Ota$^{1}$} 
\author{Y.~Kotani$^{1}$}
\altaffiliation{Present address: Japan Synchrotron Radiation Research Institute (JASRI/SPring-8), Sayo, Hyogo 679-5198, Japan}
\author{T.~Shimojima$^{1}$} 
\altaffiliation{Present address: Department of Applied Physics, University of Tokyo, Tokyo 113-8656, Japan}
\author{T.~Kiss$^{1}$}
\altaffiliation{Present address: Graduate School of Engineering Science, Osaka University, Osaka 560-8531, Japan}
\author{S.~Watanabe$^{2}$}
\author{C.~-T.~Chen$^{3}$} 
\author{S.~Niitaka$^{4,5}$} 
\author{T.~Hanaguri$^{4,5}$} 
\author{H.~Takagi$^{4,5}$}
\author{A.~Chainani$^{6,7}$}
\author{S.~Shin$^{1,5,6,8}$}
\affiliation{
$^{1}$Institute for Solid State Physics (ISSP), University of Tokyo, Kashiwa, Chiba 277-8581, Japan\\
$^{2}$Research Institute for Science and Technology, Tokyo University of Science, Chiba 278-8510, Japan\\
$^{3}$Beijing Center for Crystal R\&D, Chinese Academy of Science (CAS), Zhongguancun, Beijing 100190, China\\
$^{4}$RIKEN Advanced Science Institute, 2-1, Hirosawa, Wako, Saitama 351-0198, Japan \\
$^{5}$TRIP, JST, Chiyoda-ku, Tokyo 102-0075, Japan\\
$^{6}$RIKEN SPring-8 Center, Sayo-gun, Hyogo 679-5148, Japan\\
$^{7}$Department of Physics, Tohoku University, Aramaki, Aoba-ku, Sendai 980-8578 Japan\\
$^{8}$CREST, JST, Chiyoda-ku, Tokyo 102-0075, Japan
}

\maketitle

\subsection*{Band-structure calculations and dominant orbital characters}

Figure S1(a) shows band dispersions calculated by the Wien2k code for the pure FeTe. The lattice parameters were taken from those obtained by the powder neutron diffraction measurements~\cite{Li2009PRB} as in the previous report by Miyake {\it et al.}~\cite{Miyake2010JPSJ}. We confirmed that the obtained band dispersions are in accord with those by Miyake {\it et al}. We deduced the dominant orbital contribution of each band as indicated by different colors. Figure S1(b) shows the enlarged band dispersions in the vicinity of $E_F$ along the measured $X$-$\Gamma$-$X$ line. The 28th, 29th, and 30th bands show hole-like dispersions and cross the $E_F$, whereas the 31st band shows a electron-like dispersion just above $E_F$ around the $\Gamma$ point. 


\subsection*{Orbital characters from polarization dependent measurements}

Figure S2(a) shows the experimental configuration of the laser ARPES measurements for this study. In this configuration, $xy$ and $yz$ orbitals can be measured only for the $p$ polarization, whereas the $x^2-y^2$, $z^2$, and $xz$ orbitals can be measured for both the polarizations from the parity selection rule~\cite{Damascelli2003RMP}. By taking account of the parity of each $d$ orbital and orbital characters obtained from the band-structure calculation, we assigned the dominant orbital characters of the observed three hole bands 
as well as the electron band just above $E_F$, 
as shown in Fig. S2(c).

\subsection*{Determination of the band dispersions above $E_F$ from three different methods}

We used three different methods to determine the band dispersions above $E_F$ at 25 K as described in the following. Each method has its own advantages and disadvantages. However, the three methods provide consistent band dispersions, and we can safely conclude that an electron band exists just above $E_F$ at the $\Gamma$ point. We could then determine the positions of the bottom of the electron band and the top of $x^2-y^2$ hole band.

\subsubsection*{Second derivative spectra with respect to energy}
Figure S3(a) shows a second derivative map with respect to energy obtained from the map shown in Fig. 1(a). The open circles indicate the peak positions. Figure S3(b) is obtained by first dividing the intensity map by the Fermi-Dirac (FD) function at $T$ = 25 K broadened with the experimental energy resolution, and then taking the second derivative.  

\subsubsection*{Fitting to EDCs}
Figure S4 shows the results of fitting to several EDC cuts along the $X$-$\Gamma$-$X$ direction without dividing by the FD function. The solid lines indicate the fitting results. The fitting functions were obtained by first multiplying the FD function to the three Lorentzians, and then taking the convolution with the Gaussian corresponding to the experimental energy resolution. 
 
\subsubsection*{Fitting to MDCs}
Figures S5(a) and S5(b) show the results of fitting to several cuts of MDCs at (a) 25 K and (b) 2.5 K, respectively, after dividing by the FD function convoluted with the Gaussian. We note that dividing by the FD function does not affect the lineshape of MDCs. The solid lines and vertical bars indicate the fitting functions and their peak positions. The fitting was performed in the region of $k\ge0$ with the symmetrized Lorentzians to avoid matrix element effects, i.e., the MDC fitting function $I(\omega)$ is given by $I(\omega) = \sum_{i}I_{i}(k) + \sum_{i}I_{i}(-k)$, where $I_{i}(k)$ is a component Lorentzian. 

\subsection*{Band dispersions determined by various methods}

Figures S6(a) and S6(b) show the $E$-$k$ map measured at (a) 25 K and (b) 2.5 K, respectively. The open circles, rectangles, and triangles indicate the band dispersions determined by the second derivative spectra, fitting to the EDCs, and fitting to the MDCs. As mentioned above, each of these methods has advantages and disadvantages. Because the peak positions of the second derivative spectra are located where the gradient of the spectra shows large changes, it can detect the lower-energy side of the dispersion around the $\Gamma$ point. The fitting to the EDCs is the most appropriate to determine the band dispersion where the gradient of the dispersion is small, but it is difficult to determine the dispersions above $E_F$ without ambiguity. On the other hand, there is no ambiguity for the peak positions of MDCs even for those above $E_F$, but it is difficult to determine the dispersions around the top and bottom of the bands from MDC fits. However, the combination of three methods allows us to conclude that the top of the hole band is located at 6-7 meV above $E_F$ and an electron band exists just above $E_F$ for the dispersions at 25 K. At 2.5 K, we can recognize the superconducting coherence peaks separately at $k$ = $k_F$ and the $\Gamma$ point. 

\subsection*{Fitting to EDCs at higher temperatures}
We performed measurements with another sample at higher temperatures of 35 K and 50 K in addition to 25 K. Figures S7(a)-(c) show the FD-divided EDCs at the $\Gamma$ point for these temperatures. The solid lines indicate the fitting functions which are the same as those used in Fig. 1(c), and the dashed lines are the component Lorentzians. The peak positions are consistent with data shown in Fig. 1, although the relative intensities of the two peaks are different. Figure S7(d) shows the FD-divided EDCs along the $\Gamma$-$X$ line at 50 K and the corresponding intensity plot is shown in Fig. S7(e). The rectangles indicate the band dispersions deduced from the fitting to the EDCs  in the same way as Fig. 1(b). The solid lines are the same as those in Fig. 1(b) and extended up to 20 meV above $E_F$. It confirms that the band dispersions at 50 K are consistent with the band dispersions at 25 K. 

\subsection*{Polarization dependent spectra below $T_c$ and above $T_c$}

Figures S8 and S9 show the polarization-dependent intensity map and second derivative map measured below $T_c$ (= 2.5 K) and above $T_c$ (= 25 K), respectively. They were measured with right circular, left circular, $s$-, and $p$-polarizations. 
As described above in the section of {\bf ``Orbital characters from polarization dependent measurements''}, the orbitals with the odd parity can be 
measured with the $p$-polarization. In the second derivative spectra below $T_c$, a clear feature at $\Gamma$ point can be seen only for the $p$-polarization just below $E_F$, while the structure for $k_F$ crossings away from $\Gamma$ point can be seen for the both polarizations. This indicates a difference in orbital characters between the $x^2-y^2$ hole band and the electron band just above $E_F$. The right and left circular polarizations were used to avoid the selection rules for bands of particular symmetry which arise in photoemission with linear polarization. This ensures we have measured all the band dispersions with minimal matrix element effects. 

\subsection*{Dispersion of the superconducting coherence peak around the $\Gamma$ point}

The superconducting coherence peaks and BQP dispersions shown in Figs. 2(b) and 2(g) were enlarged in Fig. S10. These plots clearly shows that the coherence peaks at $k$ $\sim$ $\Gamma$ is originated from the electron band just above $E_F$. The BQP disperion originated from the $x^2-y^2$ hole band is almost flat  around $k$ = $\pm$0.1 {\AA}$^{-1}$ and shows an indication of a bending-back behaviour, while the curvature of the dispersion around $k$ $\sim$ $\Gamma$ is very different from the almost flat dispersion of the BQP originated from the $x^2-y^2$ hole band and corresponds a reflection of the electronic dispersion just above $E_F$ in the normal state. 

\subsection*{Temperature dependence of symmetrized EDCs and FD-divided EDCs}

The temperature dependent EDCs shown in Fig. 3 were symmetrized with respect to $E_F$ and the results are shown in Fig. S11, and they were divided by the FD functions of corresponding temperatures and the results are shown in Fig. S12. The existence of the pseudogap for the $x^2-y^2$ hole band can be clearly recognized also from both symmetrized EDCs and FD-divided EDCs as well as the temperature dependence of the SC-gap size estimated from fitting.  

\clearpage


\begin{figure*} [htbp]
\begin{center}
\includegraphics[width=12cm]{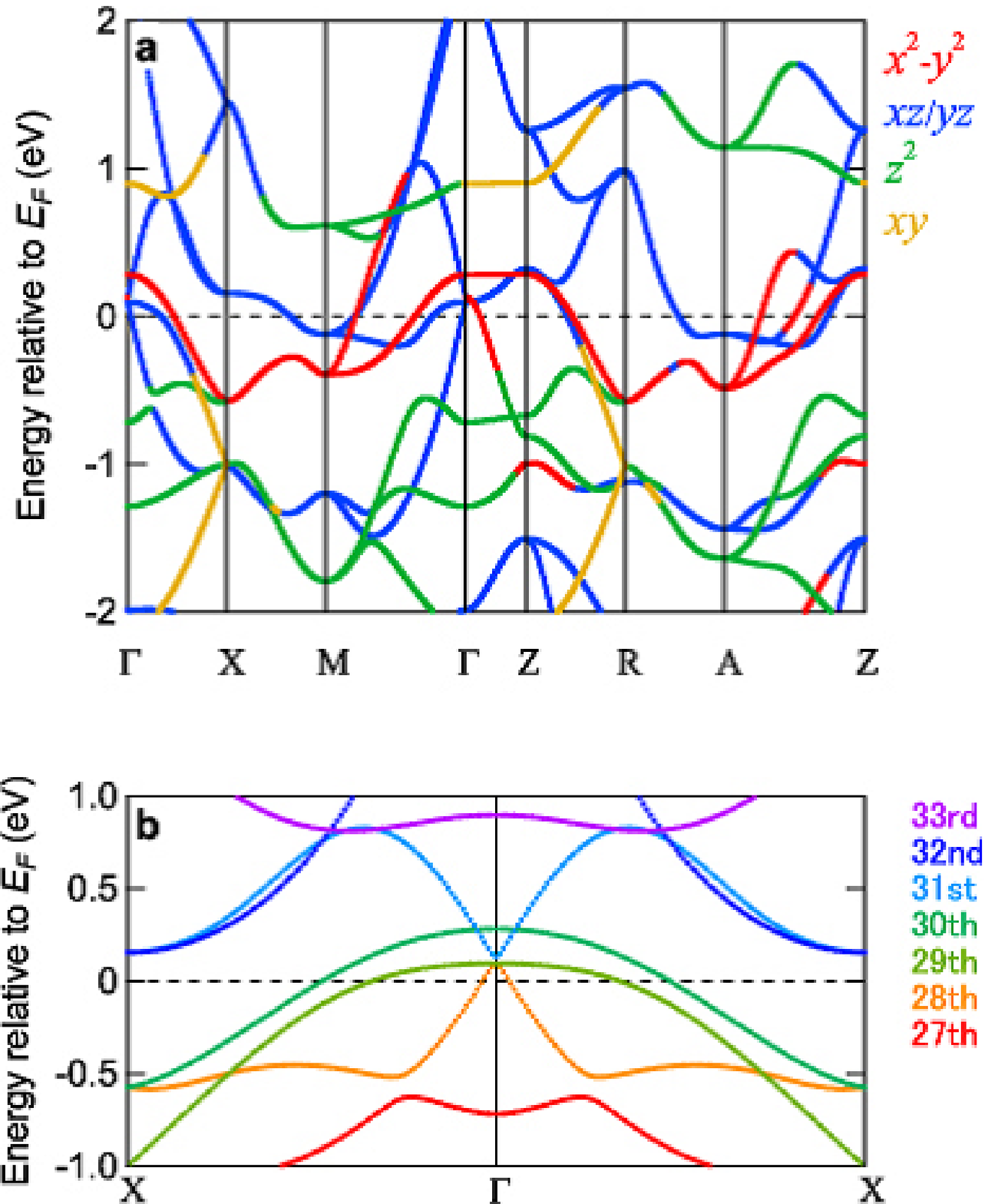} 
\end{center}
\begin{flushleft}
{\bf Fig. S1.} Band-structure calculations for the parent FeTe based on the density functional theory. ({\bf a}), Band dispersion along the high symmetric line in the Brillouin zone calculated by Wien2k code. The dominant orbital character of each band is indicated by different colors. The bands near the $E_F$ are mainly composed of $x^2-y^2$ and $xz/yz$ orbitals. ({\bf b}), Band dispersion near the $E_F$ is along the $X$-$\Gamma$-$X$ line. The 28th, 29th, and 30th bands show hole-like dispersions and cross the $E_F$, whereas the 31st band shows a electron-like dispersion just above $E_F$ around the $\Gamma$ point. The dispersion of the 28th and 31st bands looks like a Dirac cone.
\label{FigS1} 
\end{flushleft}
\end{figure*}
  
\begin{figure*} [htbp]
\begin{center}
\includegraphics[width=15cm]{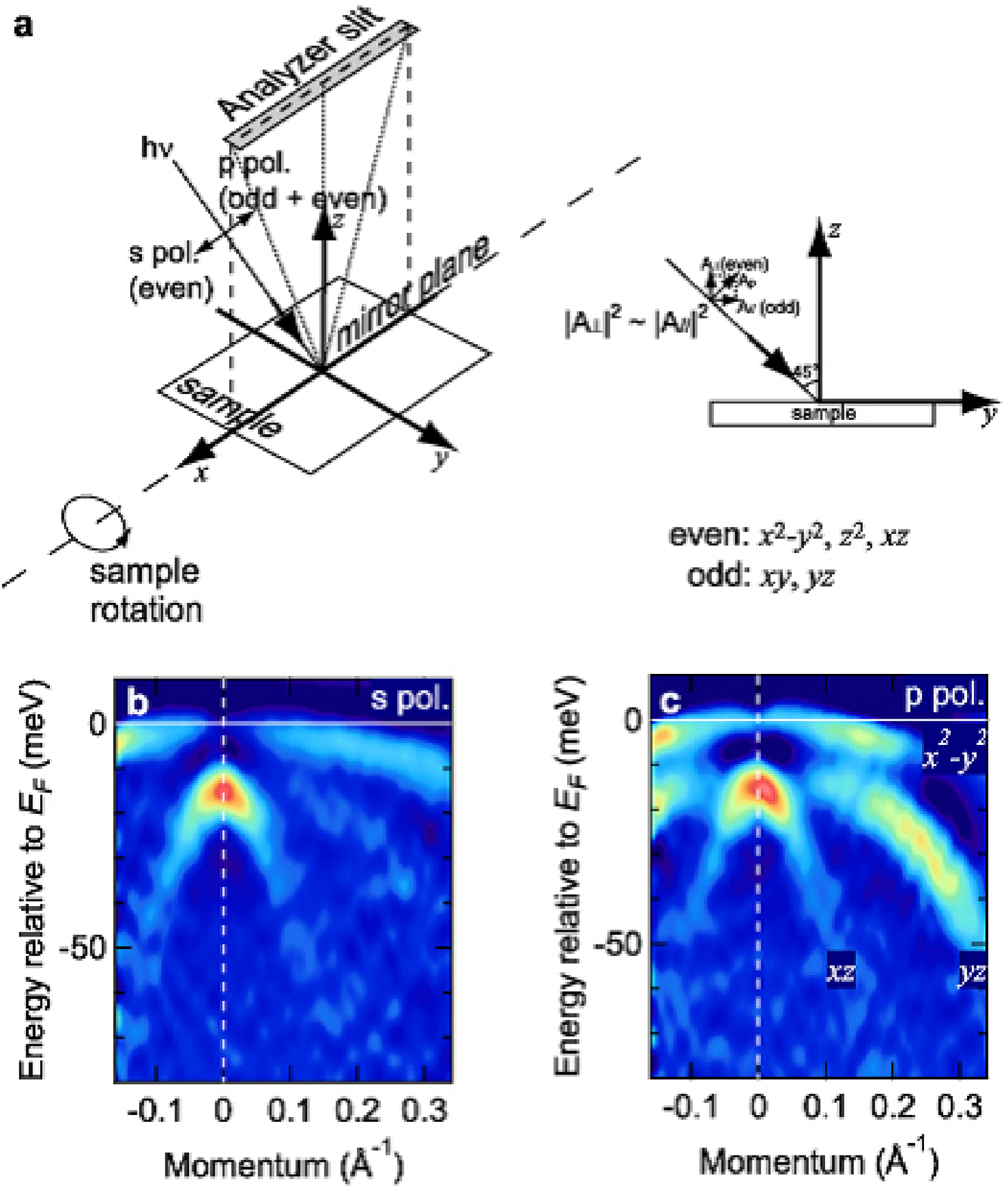}
\end{center}
\begin{flushleft}
{\bf Fig. S2.} Experimental configuration and linear-polarization dependence of ARPES intensity. ({\bf a}), Experimental configuration and parity of each $d$ orbital with respect to the mirror plane including the analyzer slit. ARPES intensity plotted as a function of momentum and energy measured at 25 K with ({\bf b}) $s$- and  ({\bf c}) $p$-polarizations, respectively. Dominant $d$ orbital for each band is indicated in (c).    
\label{FigS2} 
\end{flushleft}
\end{figure*}

\begin{figure*} [h]
\begin{center}
\includegraphics[width=15cm]{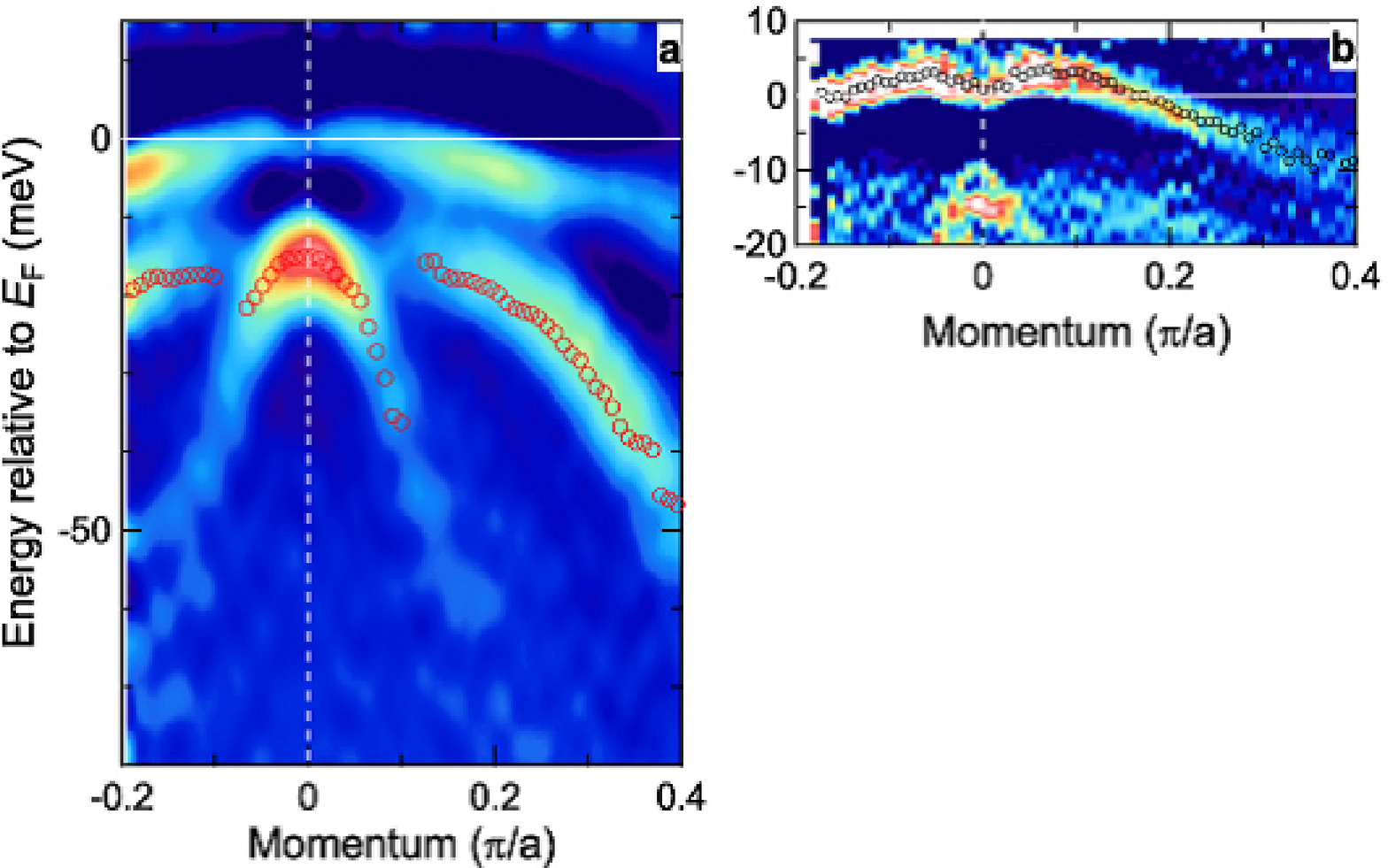}
\end{center}
\begin{flushleft}
{\bf Fig. S3.} Band dispersions from the second derivative spectra ({\bf a}), Second derivative map with respect to energy. ({\bf b}), Second derivative map after dividing by the FD function convoluted with a Gaussian of the experimental resolution. The open circles indicate band dispersions deduced from the peak positions of each map.
\label{FigS3} 
\end{flushleft}
\end{figure*}

\begin{figure*} [h]
\begin{center}
\includegraphics[width=8cm]{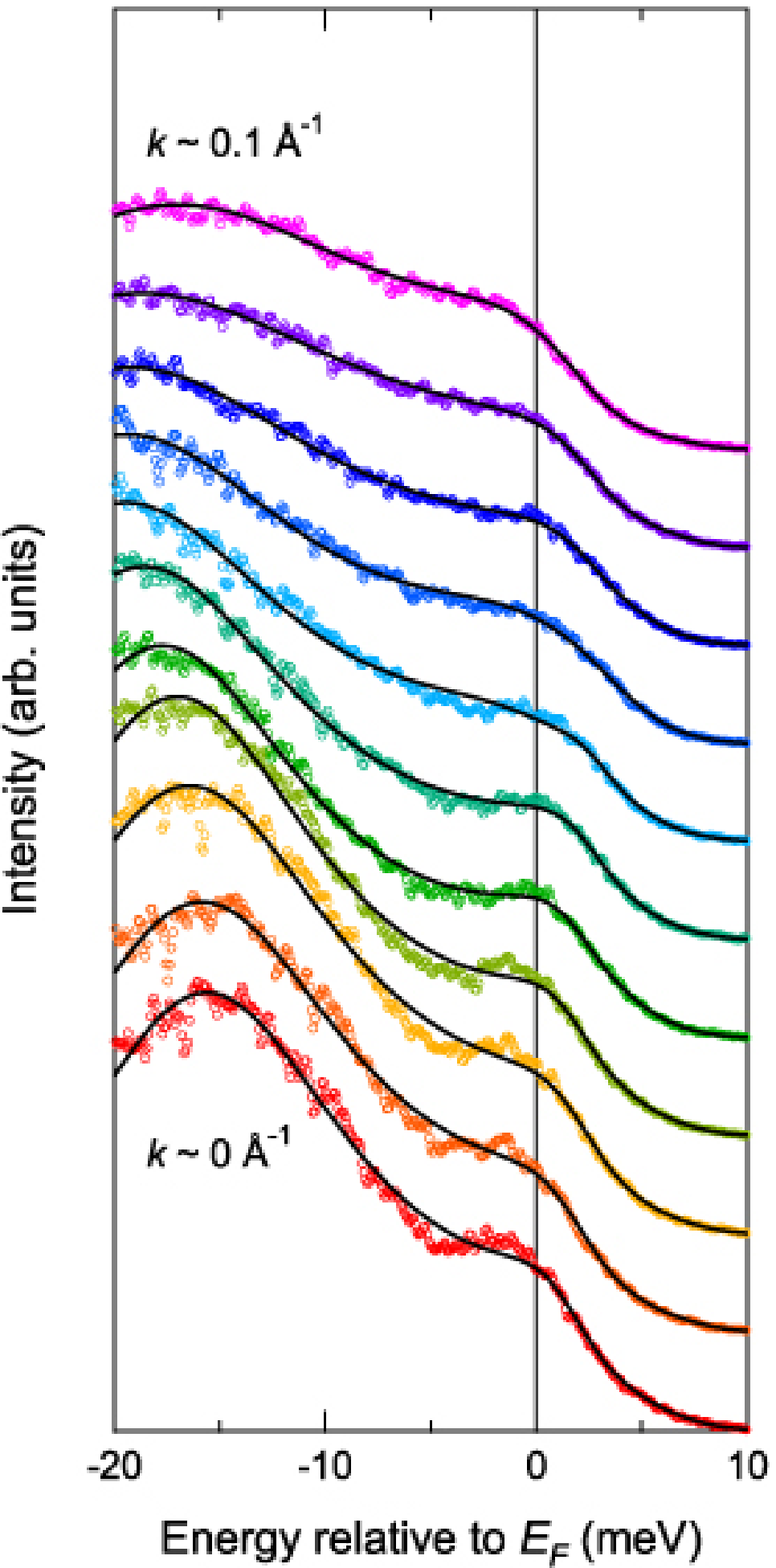}
\end{center}
\begin{flushleft}
{\bf Fig. S4.} Fitting to several cuts of EDCs along the $\Gamma$-$X$ line without dividing by the FD function. The solid lines indicate the fitting results. 
\label{FigS4}
\end{flushleft}
\end{figure*}

\begin{figure*} [h]
\begin{center}
\includegraphics[width=15cm]{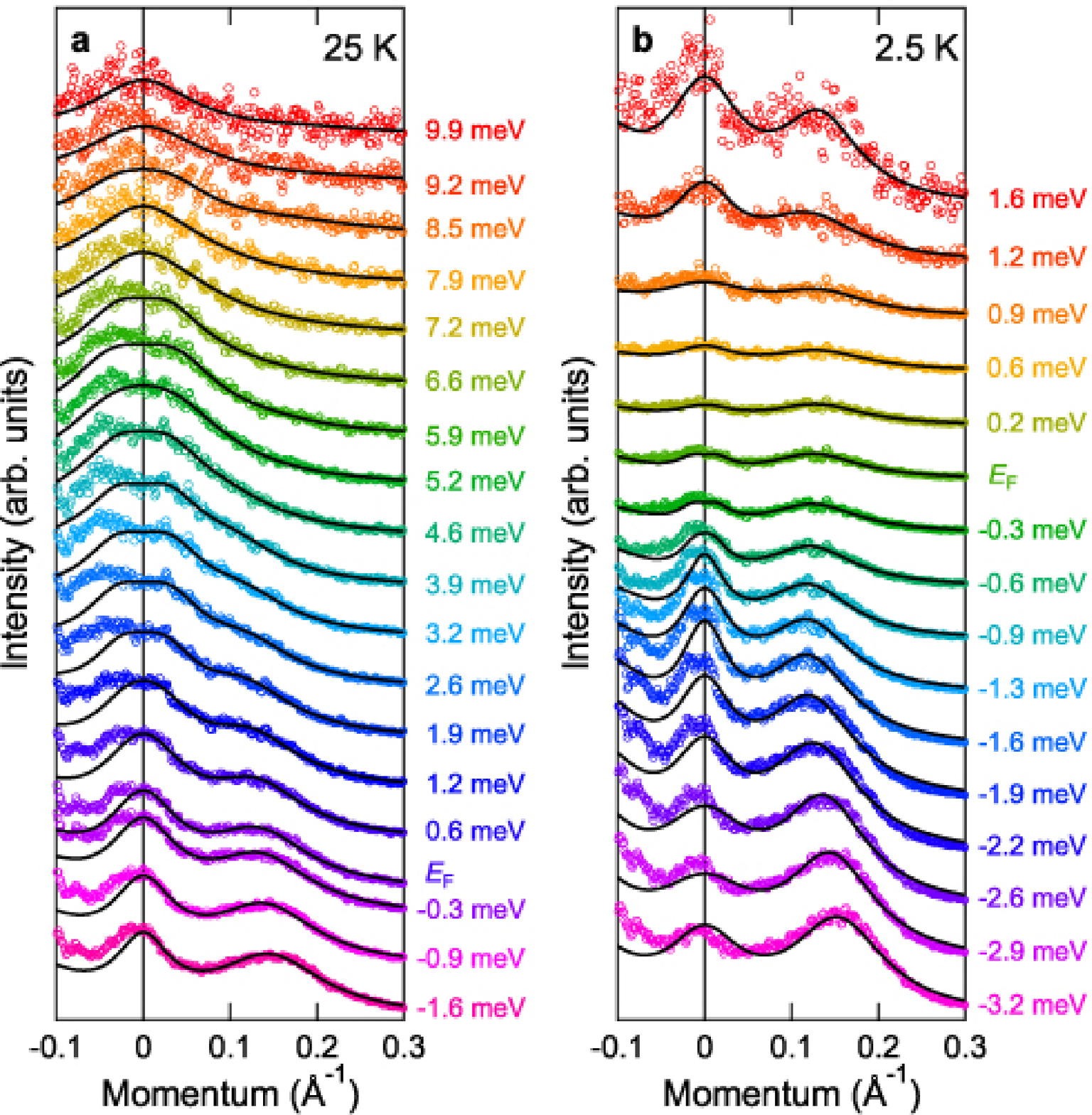}
\end{center}
\begin{flushleft}
{\bf Fig. S5.} Fits to several cuts of MDCs along the $\Gamma$-$X$ line. {\bf a} MDCs at 25 K. {\bf b} MDCs at 2.5 K.
\label{FigS5} 
\end{flushleft}
\end{figure*}
  
\begin{figure*} [h]
\begin{center}
\includegraphics[width=12cm]{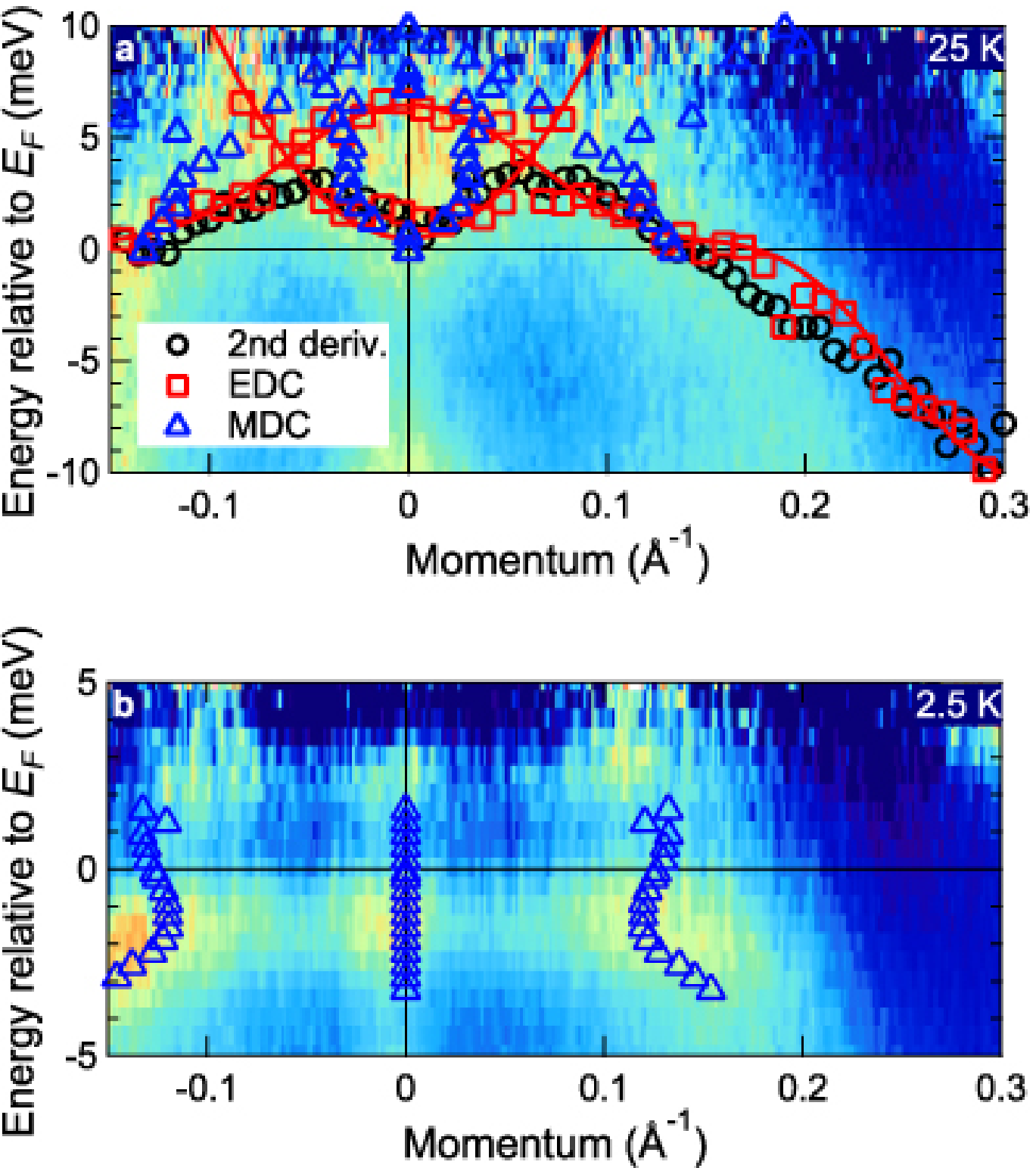}
\end{center}
\begin{flushleft}
{\bf Fig. S6.} Band dispersions determined by three methods. ({\bf a} and {\bf b}), $E$-$k$ map measured at (a) 25 K and (b) 2.5K.  Black circles, red rectangles, blue triangles are determined by the second derivative spectra, fitting to the EDCs, and fitting to the MDCs, respectively.
\label{FigS6} 
\end{flushleft}
\end{figure*}

\begin{figure*} [h]
\begin{center}
\includegraphics[width=15cm]{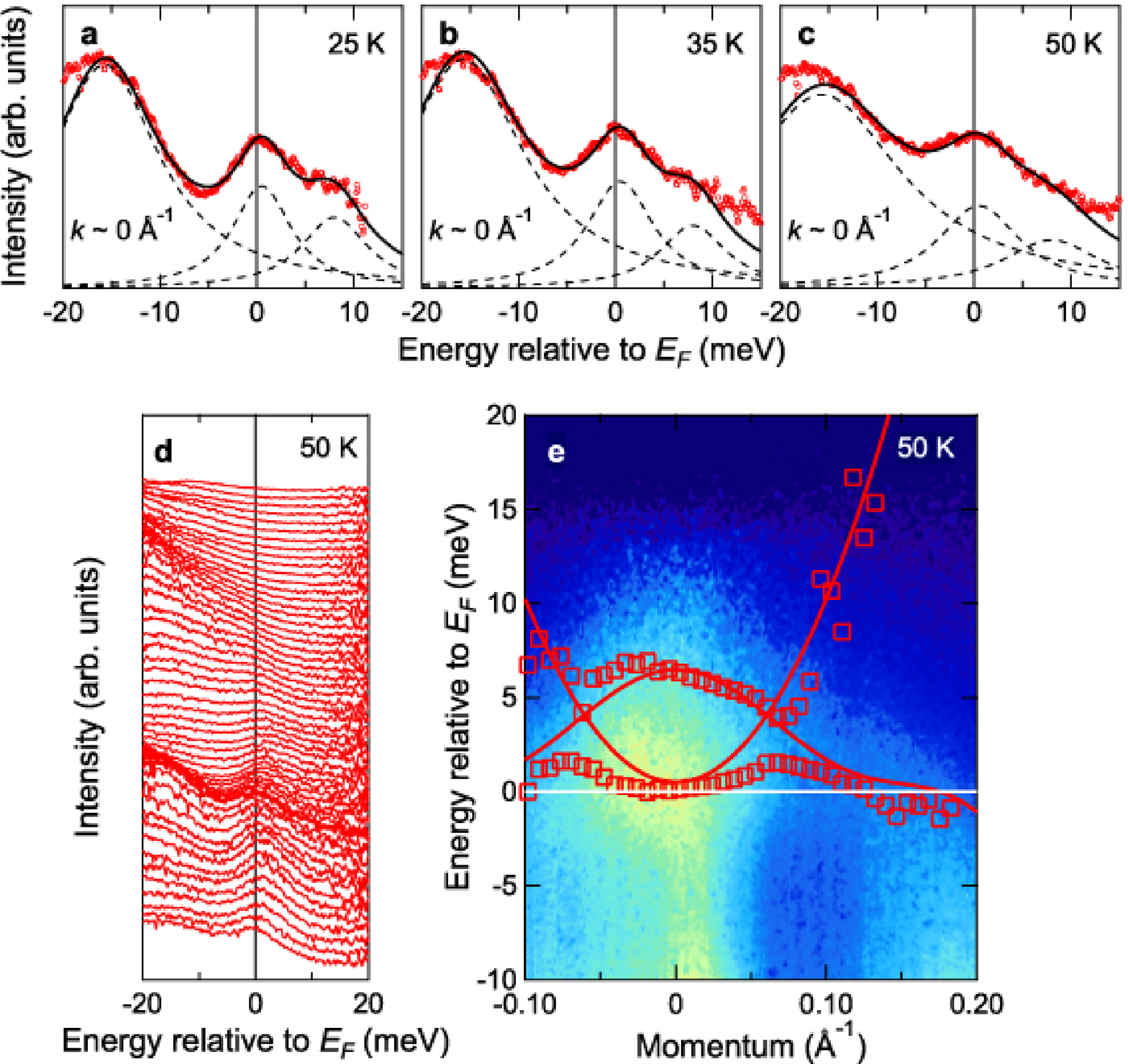}
\end{center}
\begin{flushleft}
{\bf Fig. S7.} Fitting to FD-divided EDCs at higher temperatures. ({\bf a}-{\bf c}), FD-divided EDCs at the $\Gamma$ point measured at 25 K (a), 35 K (b), and 50 K (c) for another sample. The solid and dashed lines indicate the fitting functions and component Lorentzians. ({\bf d}) FD-divided EDCs along the $\Gamma$-$X$ line at 50 K. ({\bf e}) Intensity plot of the ARPES spectra at 50 K. The rectangles indicate the band dispersions deduced from the fitting to the EDCs in the same way as Fig. 1(b). The solid lines are the same as those in Fig. 1(b), thereby confirming that the band dispersions at 50 K are consistent with the band dispersions at 25 K.
\label{FigS7} 
\end{flushleft}
\end{figure*}
  
\begin{figure*} [h]
\begin{center}
\includegraphics[width=15cm]{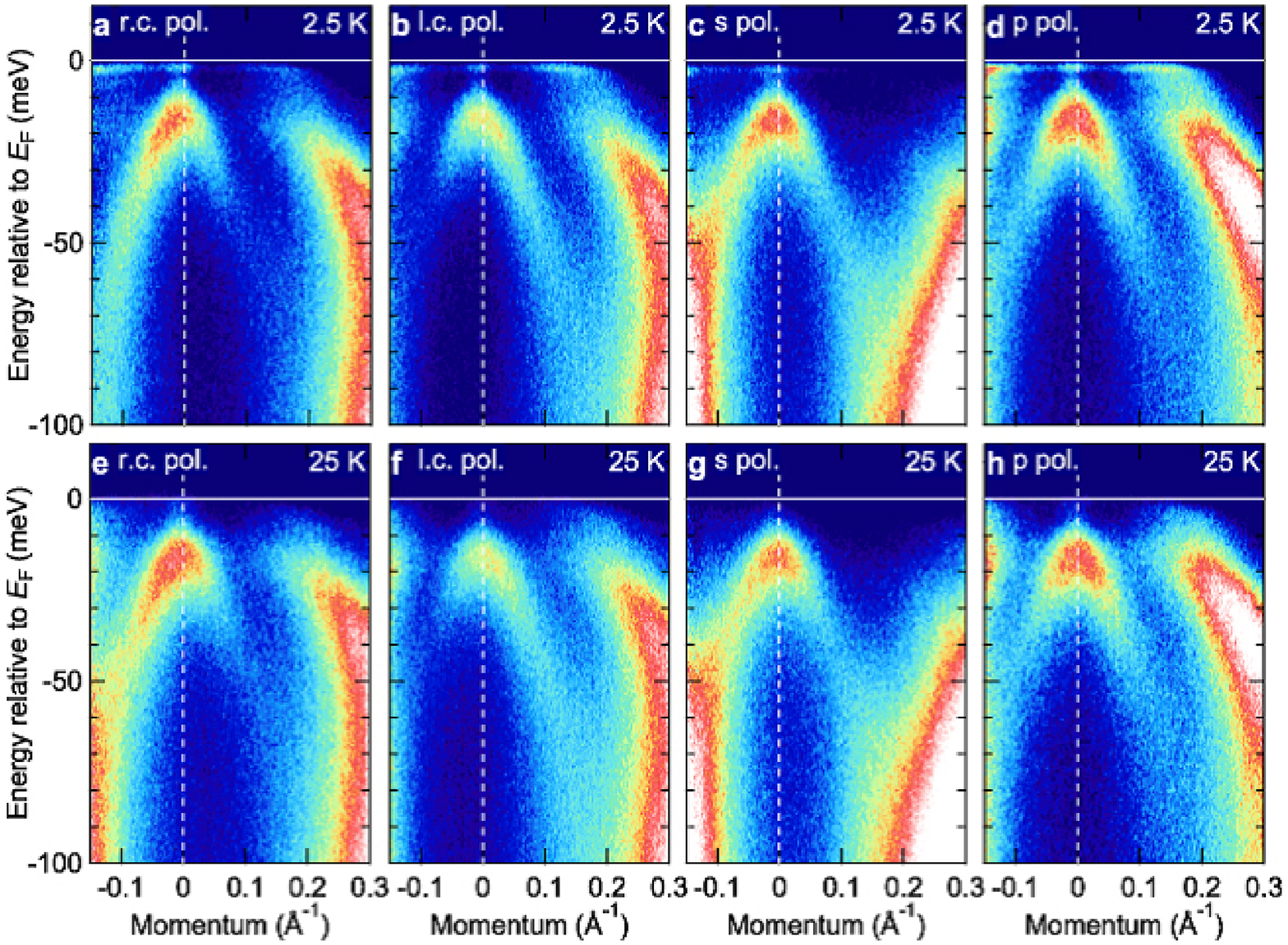}
\end{center}
\begin{flushleft}
{\bf Fig. S8.} Polarization-dependent intensity map below $T_c$ (= 2.5 K) and above $T_c$ (= 25 K). ({\bf a}-{\bf d}) Intensity map at 2.5 K measured with (a) right circular, (b) left circular, (c) $s$-, and (d) $p$-polarizations, respectively. ({\bf e}-{\bf h}) Intensity map at 25 K measured with (e) right circular, (f) left circular, (g) $s$-, and (h) $p$-polarizations, respectively.
\label{FigS8} 
\end{flushleft}
\end{figure*}
  
\begin{figure*} [h]
\begin{center}
\includegraphics[width=15cm]{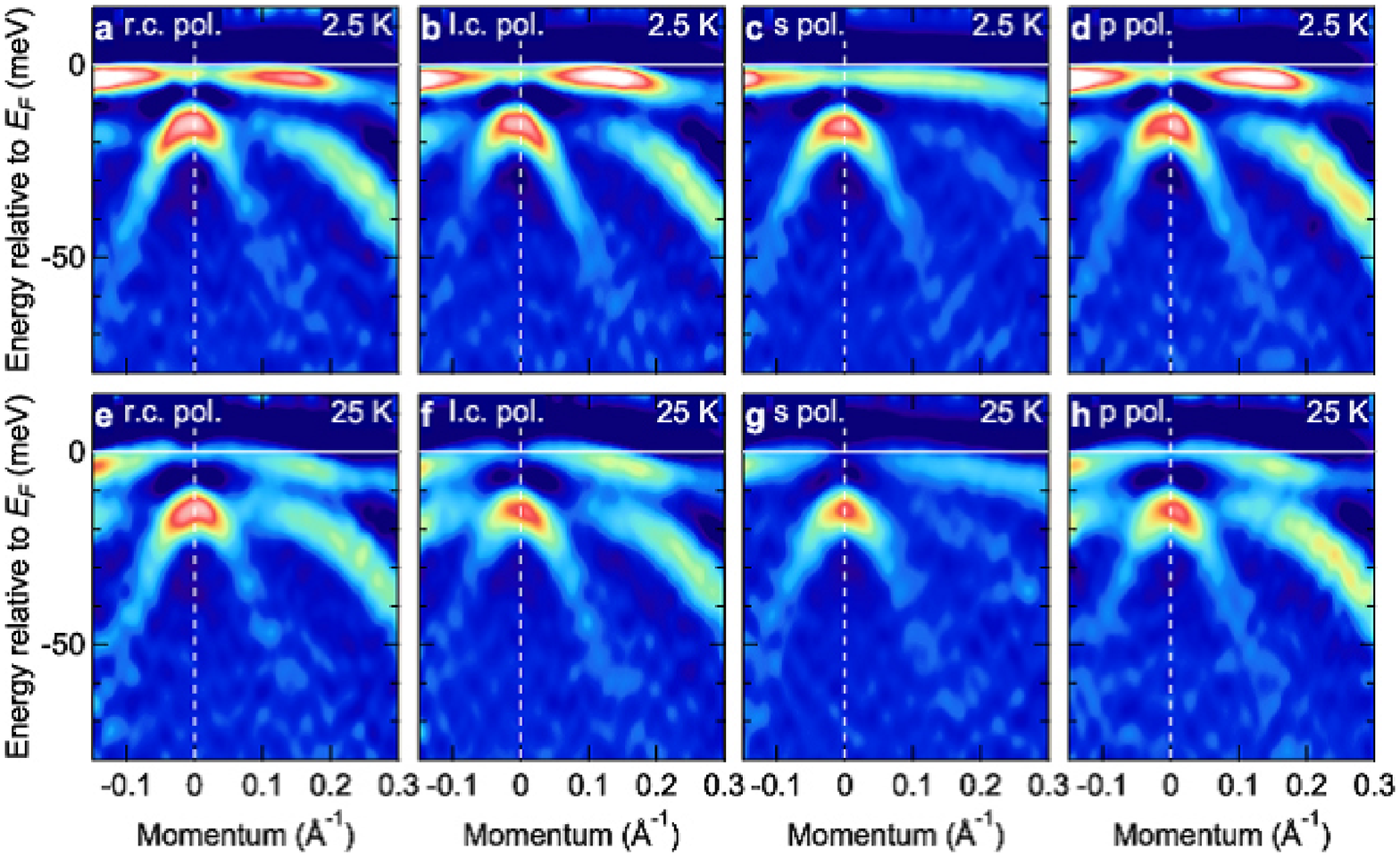}
\end{center}
\begin{flushleft}
{\bf Fig. S9.} Polarization-dependent second derivative map with respect to energy below $T_c$ (= 2.5 K) and above $T_c$ (= 25 K). ({\bf a}-{\bf d}) Intensity map at 2.5 K measured with (a) right circular, (b) left circular, (c) $s$-, and (d) $p$-polarizations, respectively. ({\bf e}-{\bf h}) Intensity map at 25 K measured with (e) right circular, (f) left circular, (g) $s$-, and (h) $p$-polarizations, respectively.
\label{FigS9} 
\end{flushleft}
\end{figure*}
  
\begin{figure*} [h]
\begin{center}
\includegraphics[width=14cm]{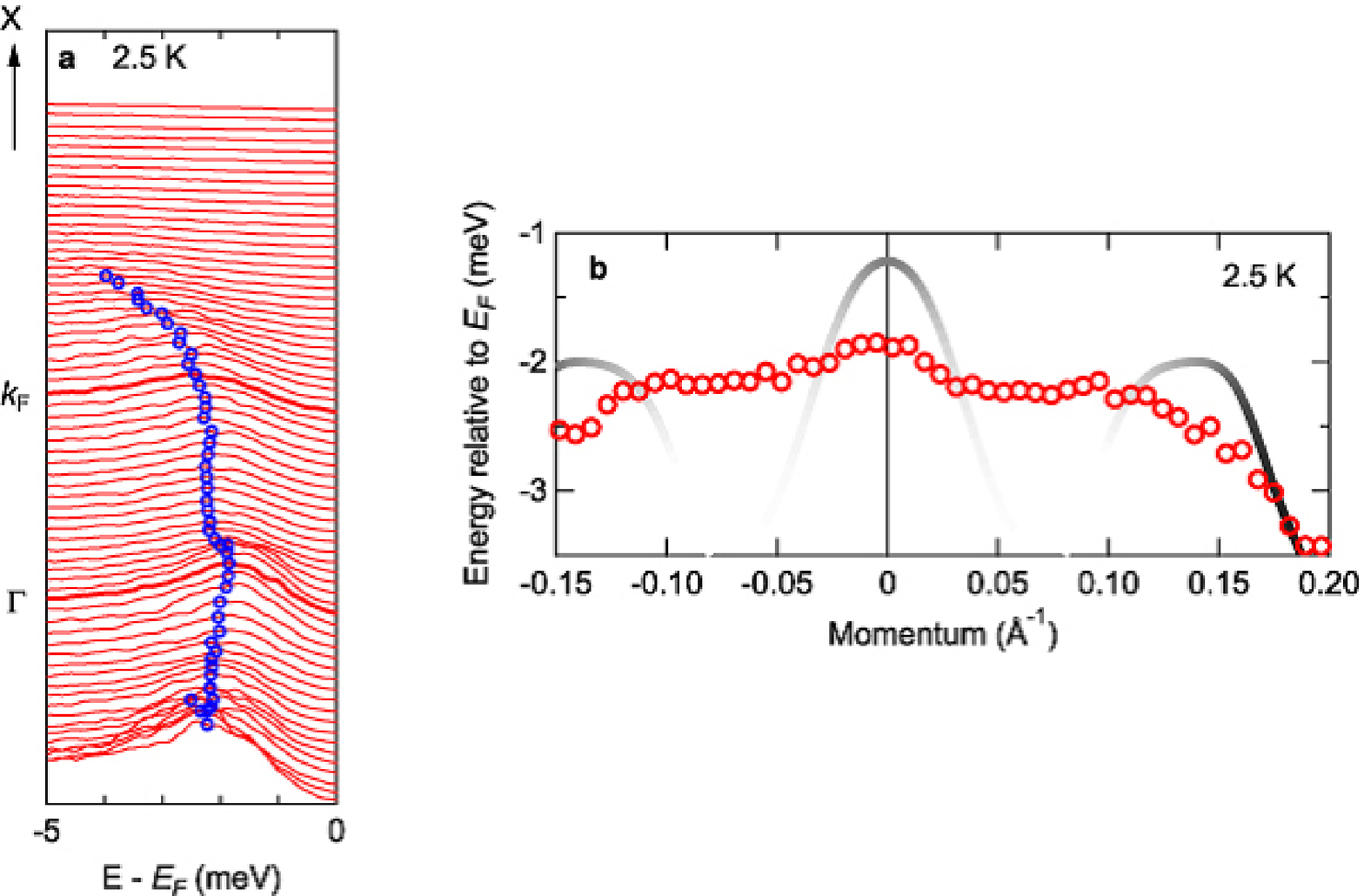}
\end{center}
\begin{flushleft}
{\bf Fig. S10.} The superconducting coherence peaks and BQP dispersions shown in Figs. 2(b) and 2(g) are plotted in an enlarged scale. These plots clearly show that the coherence peaks at $k$ $\sim$ $\Gamma$ is originated from the electron band just above $E_F$.\label{FigS10} 
\end{flushleft}
\end{figure*}

\begin{figure*} [h]
\begin{center}
\includegraphics[width=14cm]{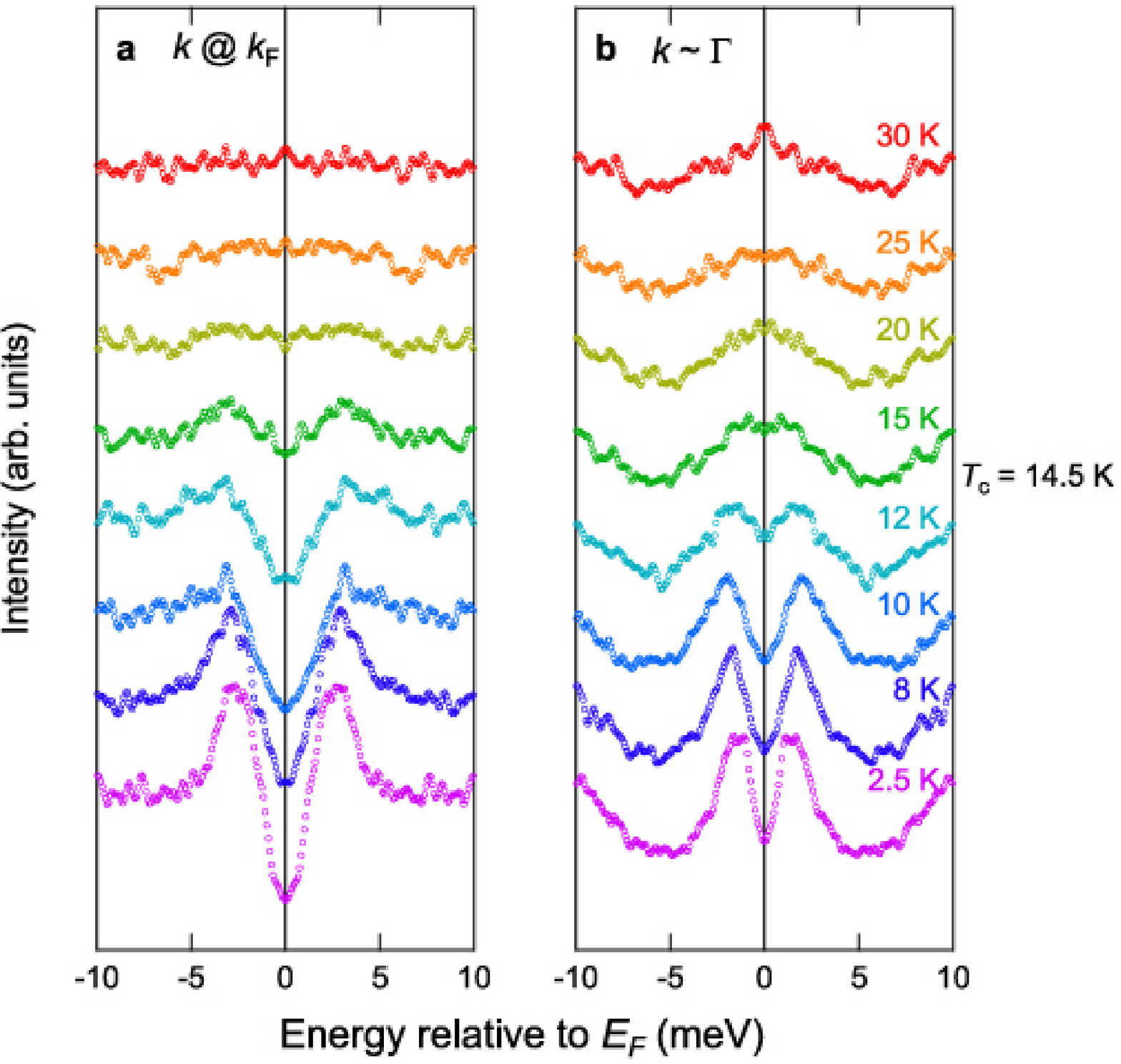}
\end{center}
\begin{flushleft}
{\bf Fig. S11.} Temperature dependence of symmetrized EDCs at $k$ = $k_F$ and $k$ $\sim$ $\Gamma$. Temperature dependence of symmetrized EDCs at ({\bf a}) $k$ = $k_F$ of $x^2-y^2$ hole-like band and ({\bf b}) $k$ $\sim$ $\Gamma$ (bottom of the electron-like band), respectively.\label{FigS11} 
\end{flushleft}
\end{figure*}
  
\begin{figure*} [h]
\begin{center}
\includegraphics[width=14cm]{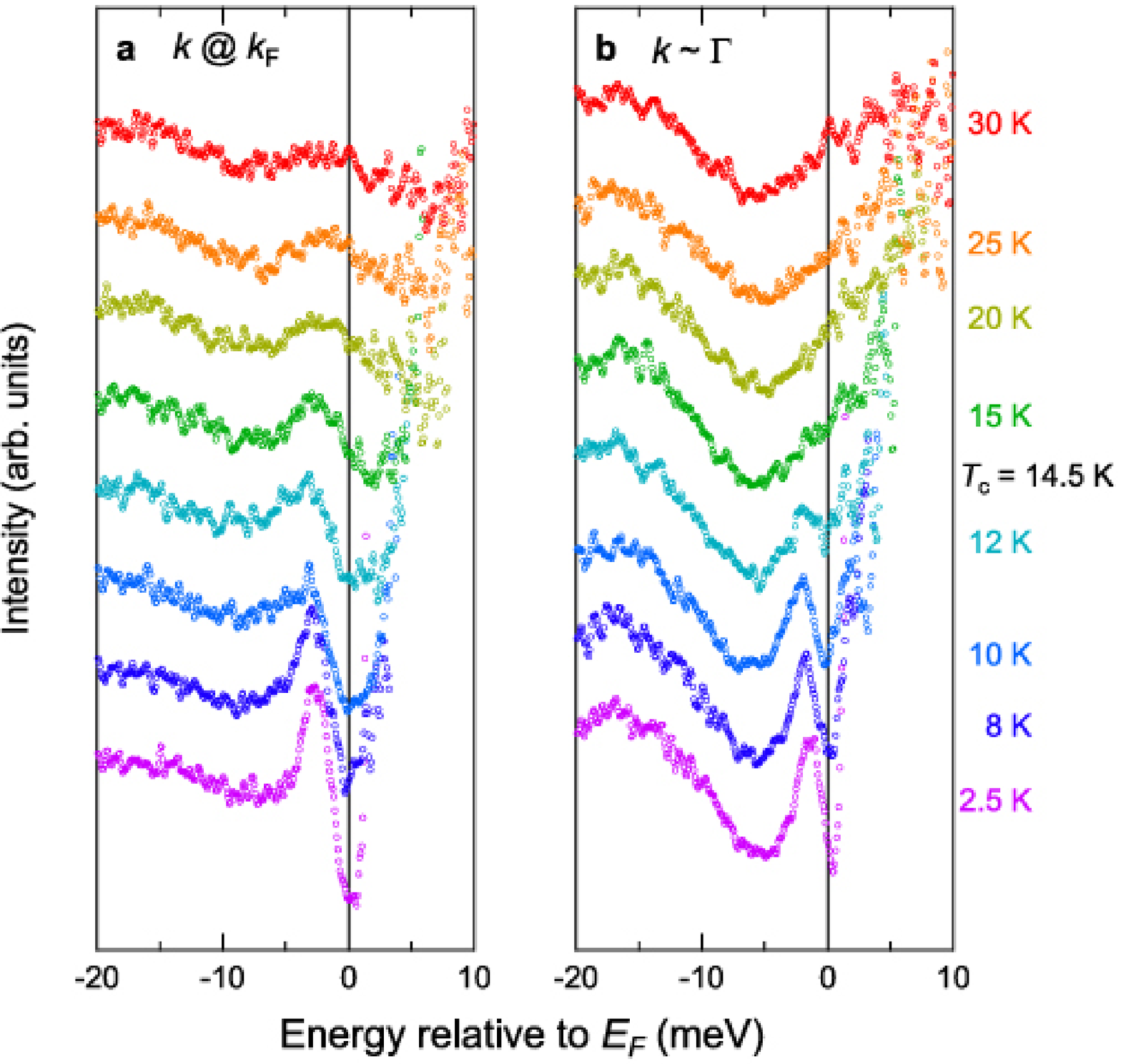}
\end{center}
\begin{flushleft}
{\bf Fig. S12.} Temperature dependence of FD-divided EDCs at $k$ = $k_F$ and $k$ $\sim$ $\Gamma$. Temperature dependence of FD-divided EDCs at ({\bf a}) $k$ = $k_F$ of $x^2-y^2$ hole-like band and ({\bf b}) $k$ $\sim$ $\Gamma$ (bottom of the electron-like band), respectively.\label{FigS12} 
\end{flushleft}
\end{figure*}


